\documentclass[onecolumn]{svjour2}                    
\smartqed  
\usepackage{graphicx}
\usepackage{dcolumn}
\usepackage{bm}
\usepackage{amssymb}%
\usepackage{mathptmx}      

\journalname{Foundations of Physics}
\begin{document}

\title{The Logic of Identity: Distinguishability and Indistinguishability in Classical and Quantum Physics\footnote{Dedicated to the memory of N.G.\ van Kampen (1921-2013)}}
\titlerunning{Logic of Identity}

\author{Dennis Dieks}

\institute{D. Dieks \at
              Institute for the History and
Foundations of Science, Utrecht University\\ P.O.Box 80.010, 3508
TA Utrecht, The Netherlands \\
              \email{d.dieks@uu.nl} }

\date{Received: date / Accepted: date}

\maketitle

\begin{abstract}
The suggestion that particles of the same kind may be indistinguishable in a fundamental sense, even so that challenges to traditional notions of individuality and identity may arise, has first come up in the context of classical statistical mechanics. In particular, the Gibbs paradox has sometimes been interpreted as a sign of the untenability of the classical concept of a particle and as a premonition that quantum theory is needed. This idea of a `quantum connection' stubbornly persists in the literature, even though it has also been criticized frequently. Here we shall argue that although this criticism is justified, the proposed alternative solutions have often been wrong and have not put the paradox in its right perspective. In fact, the Gibbs paradox is unrelated to fundamental issues of particle identity; only distinguishability in a pragmatic sense plays a role (in this we develop ideas of van Kampen \cite{vankampen}), and in principle the paradox always is there as long as the concept of a particle applies at all. In line with this we show that the paradox survives even in quantum mechanics, in spite of the quantum mechanical (anti-)symmetrization postulates.
\end{abstract}

\section{Introduction}
Questions about identity and distinguishability in modern physics have their origin in thermodynamics and statistical physics.
In particular, the notorious Gibbs paradox raises the question of exactly when two gases can be considered as being of the same kind, and what consequences this sameness has for thermodynamical properties, in particular entropy values. There is a persistent, though controversial and contested, claim in the literature that a fully satisfactory solution of the Gibbs paradox can only be achieved in quantum mechanics, via the way quantum mechanics deals with `identical particles'. Evidently, the background of this claim is the conviction that there exists a relation between quantum indistinguishability and `lack of particle identity' on the one hand, and the thermodynamical mixing with which the Gibbs paradox is concerned on the other (see for a sketch of the general background of these ideas, e.g., \cite{saundersindist}).

In this article we shall first look at the definition of the entropy of mixing, in order to emphasize the essential role of \emph{pragmatic} considerations in the Gibbs paradox (in contradistinction to fundamental considerations about particle identity). We also aim to show that existing discussions of the Gibbs paradox have usually not gone to the heart of the matter or are simply wrong. Finally, we demonstrate that the Gibbs paradox can be formulated even in the context of quantum mechanical systems consisting of `identical particles'.

\section{Identity and distinguishability in thermodynamics: the Gibbs paradox}
 Traditional thermodynamics deals with macro systems in thermal equilibrium---a paradigmatic case is a gas in a container, whose volume $V$ can be changed by means of a piston and to which heat can be supplied by contact with a heat bath. The thermodynamic state of the gas (which we here will write as a function of its pressure $P$ and temperature $T$) will be changed when we vary the volume and/or supply heat. If this is done in such a gradual way that the gas can be considered to be in equilibrium at each instant, the change is called \emph{reversible}. It is now borne out by experience that the quantity $\int_{1}^{2} dQ/T$, i.e.\ the supplied heat divided by the temperature, integrated along a reversible path from state $1$ to state $2$ in the macroscopic state space parametrized by $P$ and $T$, is independent of the path between $1$ and $2$. This means that a \emph{state function}, the \emph{entropy} $S$, can be defined such that $S(P_{2},T_{2})-S(P_{1},T_{1})= \int_{1}^{2} dQ/T$. Further, energy conservation implies $dQ-PdV=dE$.  Given that the energy $E$ of an ideal classical gas consisting of $N$ atoms is given by $E=\frac{3}{2}NkT + const$, we find $$\frac{dQ}{T}= \frac{5}{2}Nkd(\ln T) - Nkd(\ln P).$$ So the entropy of an ideal gas is determined as
\begin{equation} \label{entropy}
S(P,T)=\frac{5}{2}Nk\ln T - Nk\ln P + C.
\end{equation}
The above derivation, starting from experimental facts, does not completely fix the number $C$: $C$ does not depend on the values of $P$ and $T$ of our system, but may depend on anything else, and may in principle be taken as different when we go from one system to another, as van Kampen (\cite{vankampen}) emphasizes. Nevertheless, there exist natural choices for $C$: it is eminently reasonable and convenient to take the same value of $C$ for all systems that consist of equal amounts of the same gas, and also to take the entropy of a system that is formed by juxtaposing (without thermal contact or other interaction) two thermodynamical systems as the sum of the individual entropies, so that the entropy becomes \emph{additive} (\cite[p.\ 305]{vankampen}). This goes beyond direct empirical regularities, since these only show additivity of entropy \emph{differences}.

Combining two volumes of the same ideal gas into one volume can be done by a reversible process in which no heat is exchanged (namely, by slowly removing a partition). This, plus the just-mentioned conventions about the constants in the entropy expressions, implies the \emph{extensivity} of the entropy of ideal gases: before the removal of the partition the total entropy is double the entropy of each of the individual systems (by virtue of additivity) and this value does not change during the reversible removal of the partition, since $dQ=0$ during this process.\footnote{See (\cite{versteegh} for a discussion of the consequences of choosing conventions that do not make the entropy extensive.} So the consideration of reversible processes in which $N$ changes makes it possible to say more about the $N$-dependence of $S$ than what is already contained in (\ref{entropy}). The formula for the entropy of an ideal gas now becomes
\begin{equation}\label{entropy2}
S(P,T,N)=\frac{5}{2}Nk\ln T - Nk\ln P + cN
\end{equation}
in which the new constant $c$ does not depend on $P$, $T$ or $N$ but may still depend on other things, like the kind of gas that is being considered (\cite[p.\ 306]{vankampen}).

This derivation of the extensivity of the entropy of ideal gases highlights that in order to apply the formula $dS = \int dQ/T$ it is necessary to consider processes by means of which the number of particles can be varied in a reversible way: it would be impossible to derive an $N$-dependence of the entropy from merely studying the thermodynamical properties of closed systems. Further, as we have seen, the full dependence of the entropy on $N$ does not follow immediately from experimental results (which give us only entropy differences) but needs a choice of constants. This warns us that if we are going to seek a microscopic counterpart of the thermodynamical entropy expression, we should focus on $\triangle S$ rather than on $S$ itself. It is not to be expected that in statistical mechanics there exists a fundamental and unique justification for the absolute value of $S$, if the task of statistical mechanics is to reproduce empirical results; the freedom in $S$ that we encounter in thermodynamics should also be present in statistical mechanics.

A further comment is that obviously even the slowest removal of a partition between two ideal gases of the same kind will lead to violent processes \emph{on the micro-scale}: atoms will suddenly have more space available to them and will escape from their original volume. If the two gases are not of the same chemical sort, the resulting mixing of particles from the two volumes will show itself by changes that count as thermodynamically observable and the process will count as irreversible. Such an irreversible process will increase the entropy even if $dQ = 0$. But if the two gases are chemically speaking the same, the mixing will not be detectable by looking at the usual thermodynamical quantities. This is so because in thermodynamics we restrict ourselves to the consideration of coarse-grained macroscopic quantities, and this entitles us to describe the mixing of two volumes of gases of the same kind, with equal $P$ and $T$, as \emph{reversible} with no increase in entropy. But if we think of what happens in terms of the motions of individual atoms or molecules, the two processes (irreversible and reversible mixing) are completely similar. In other words, the qualification of the mixing process as irreversible or reversible, and the verdict that the entropy does or does not change, possesses a pragmatic dimension. It depends on what we accept as legitimate methods of discrimination (\cite{vankampen}): chemical differences lead to acknowledged thermodynamical entropy differences in a process of mixing, whereas mere differences in where particles come from do not.

Indeed, in the case of two chemically different ideal gases, the description on the level of thermodynamics permits that the gases can be effectively distinguished: it is assumed, as part of the thermodynamical framework, that semi-permeable membranes exist that are transparent to one gas but stop the particles of the other. A reversible mixing process can now be defined via the familiar procedure in which such membranes are slowly moved, while the gas whose atoms or molecules cannot pass is exerting a pressure on the membrane. Calculation of the work that is done and the heat that has to flow in in order to keep the temperature constant leads to the familiar result that the mixing results in an increase in entropy. The value of this `entropy of mixing' is independent of the difference between the two gases, and remains the same when (in thought) we make the gases more and more similar. For two equal initial volumes $V$ of ideal gases, each with the same $P$ and $T$ and each with $N$ atoms, which are mixed so that they are both contained in one volume $2V$, the increase of entropy is given by
\begin{equation} \label{mixing}
S_{mix}=2kN\ln 2.
\end{equation}
That (\ref{mixing}) does not gradually vanish when the gases become more and more similar is one form of the Gibbs paradox.

The reason that the entropy of mixing does not diminish when two gases become more and more similar is readily understandable from the above. The derivation of Eq.\ (\ref{mixing}) as we sketched it has as its premise that the two gases can be reversibly mixed and separated by some device, for example a set of semi-permeable membranes. If this premise is granted, the outcome (\ref{mixing}) follows from calculating $\int dQ/T$ during the process. However, as soon as this premise is dropped, i.e.\ as soon as we no longer assume the availability of an effective discrimination technique, no increase of thermodynamical entropy will be found. The discontinuous transition from $2kN\ln 2$ to $0$ thus reflects the discontinuous step from considering two things as \emph{distinguishable} to considering them as \emph{indistinguishable}---there is nothing mysterious about this.

Continuing our earlier comment about the difference between irreversible and reversible, we stress again that this Gibbs discontinuity contains a pragmatic element. If we decided to follow individual particles on their trajectories, we could in principle build semi-permeable membranes that could distinguish between particles coming from the two initial volumes, even if the gases are chemically equal. With the help of such devices we could verify the presence of an entropy of mixing $S_{mix}=2kN\ln 2$ (\cite{dieks2}) even in this case. Such devices would have to make use of a kind of Maxwellian demons: little instruments that should calculate for each approaching atom or molecule whether it initially came from volume $1$ or $2$. Depending on the result of this calculation, the atom/molecule should be stopped or not. Of course, the consideration of membranes equipped with such microscopic tools or Maxwellian demons falls outside the area of competence of traditional thermodynamics. But this is exactly the point: it is our \emph{decision} to confine ourselves to the consideration of traditional thermodynamical quantities that is responsible for the disappearance of effects of mixing in the case of gases of the same chemical kind.

The discontinuous vanishing of the entropy of mixing in thermodynamics therefore does not correspond to some indistinguishability \emph{in principle}, on the level of the particles described by fundamental microscopic physics. Indeed, if we transcend the boundaries of thermodynamics it is possible, within the conceptual framework of classical physics, to recover an analogue of the entropy of mixing even when thermodynamics tells us that there is no such entropy. It follows that the presence or absence of a thermodynamical entropy of mixing does not teach us anything definitive about microscopic distinguishability and identity. These simple considerations, which in principle are well known, already indicate that attempts to connect the Gibbs paradox to issues of fundamental distinguishability and identity, perhaps relating to quantum mechanics, are misguided.

\section{Identity and distinguishability in statistical mechanics}\label{sectId}
One of Boltzmann's seminal ideas was that in statistical mechanics the thermodynamical entropy relates to the `thermodynamical probability' $W$ of a macro-state, viz.\ the number of micro-states compatible with the macro-state in question divided by the total number of micro-states. Boltzmann used a formula for entropy differences, of the form (\cite[p.\ 176]{ehrenfest})
\begin{equation} \label{boltzmann}
S_{2} - S_{1} = k\ln\frac{W_2}{W_1},
\end{equation}
in accordance with the fact that in thermodynamics only entropy differences have direct empirical significance---as we have seen, the absolute value of the entropy depends on our choice of constants. However, in the later literature Eq.\ (\ref{boltzmann}) has usually been replaced by the formula
\begin{equation} \label{boltzmann2}
S=k\ln W,
\end{equation}
with $W$ the number of micro-states. Ehrenfest and Trkal (\cite[p.\ 176]{ehrenfest}) ascribe the predominance of formula (\ref{boltzmann2}) to the influence of Planck's work. Eq.\ (\ref{boltzmann2}) has the advantage that it is simple and corresponds to a natural choice of constants. However, because it applies to a closed system, it tends to obscure the fact that also in statistical mechanics we can only learn something about the entropy's dependence on any particular quantity if we investigate processes in which the quantity in question can vary---such processes will tell us how the entropy \emph{changes} when the variable changes. In particular, $S$'s dependence on $N$ must be studied in situations in which the number of particles can vary.

If we have two equal volumes of ideal gases of the same kind, both in internal thermal equilibrium at $P$ and $T$, with a partition between them, application of Eq.\ (\ref{boltzmann2}) leads to the result that the total entropy is double the value of the individual entropies (since the total number of microstates is the product of the numbers of microstates of the individual systems, if there are no microscopic correlations between the systems). Equation (\ref{boltzmann2}) therefore embodies the standard choice of constants that makes the entropy additive. When we remove the partition, however, Eq.\ (\ref{boltzmann2}) does not produce the expected result that the combined entropy is double the value of each of the entropies of the single systems. Indeed, if the number of microstates in each of the single volumes of ideal gas at $P$ and $T$ is $W$, the number of microstates in the combined volume is not $W^2$ but rather 
\begin{equation} \label{microstates}
W^2\frac{2N!}{N!N!},
\end{equation}
in which the combinatorial factor represents the number of ways the $2N$ atoms can be distributed over the two volumes after the partition has been taken out (we assume $N$ to be very large in this calculation and similar ones, so that the total number of states can be taken as equal to the number of states of the equilibrium situation). 

The presence of this combinatorial factor comes from the fact that we can distinguish and follow in time each one of the individual particles, so that we can tell whether it originated in the left or right compartment. Before the partition was removed the total number of microstates was $W^2$, but after the removal each configuration can be realized in more ways because particles from the left can now occupy positions at the right, and \emph{vice versa}. We see why the existence of trajectories is important here: by their trajectories we can distinguish atoms from volume $1$ from those originating from volume $2$, also after the mixing. In the case of classical particles this assumption of distinguishability via trajectories is of course always satisfied. Now, if we use $S = k \ln W$ to calculate the entropy of the combined gas volumes, this gives us a value of the entropy after mixing of
\begin{equation} \label{gibbs2}
2k \ln W + k \ln \frac{2N!}{N!N!} = 2k \ln W + 2kN \ln 2,
\end{equation}
where Stirling's approximation has been used. We see that an entropy of mixing appears, just as in Eq.\ (\ref{mixing}). This leads to a variation on the original Gibbs paradox: the statistical treatment, unlike the thermodynamical approach, doesn't make the entropy of mixing vanish even when the two gases are of the same chemical kind.

A familiar strategy here is to let this unexpected entropy growth go away by inserting a factor $1/N!$ in the expression for $W$. This manoeuvre removes the entropy of mixing when we use formula (\ref{boltzmann2}), so that the entropy becomes extensive. Instead of $k \ln W$ for each of the two individual volumes before mixing we now have $k \ln W/N!$, and for the entropy of the combined system after mixing we find $$k\ln (W^2.\frac{2N!}{N!N!}.\frac{1}{2N!}),$$ which is exactly double the entropy of each of the original volumes. So introducing the additional factor $1/N!$ in all entropy formulas removes the Gibbs paradox in its second version.

\section{The factor $1/N!$}

In order to decide whether, and if so in what respect, we have come closer to a solution of the paradox, we should first of all pay attention to the objection that the division of the number of microstates by $N!$ is \emph{ad hoc} and mysterious---a criticism that has been discussed in the literature since Gibbs's days. There is a traditional reply here, going back to Gibbs himself and to Planck (see \cite{ehrenfest,vankampen,saundersindist} for references to the original literature), in which this division is defended by an appeal to the fundamental \emph{indistinguishability} of the atoms (or other particles) that are involved.\footnote{In the later literature quantum mechanics and the symmetrization postulates for states of `identical quantum particles' are frequently invoked as the final justification.}

A recent example of this line of argument is Saunders's observation that in many-particles states of particles of the same kind permutations of the particle labels do not lead to a new physical situation (\cite{saundersindist}). From the viewpoint of what can be ascertained empirically, classical particles are individuated solely by their physical properties, and it does not make any difference how we distribute indices over them. We can therefore without any loss of physical information go over to the \emph{reduced} phase space, in which all states that can be transformed into each other by permutations of indices are identified: the different assignments of particle labels do not correspond to empirical differences and can therefore be considered to represent unnecessary surplus theoretical structure. So we can get by with a phase space whose number of states is a factor $N!$ smaller than if we count all permuted states as different. According to Saunders this observation justifies the insertion of a factor $1/N!$ in the number of states, which in turn solves the Gibbs paradox.

It is certainly true that in this sense classical particles with the same intrinsic properties (mass, charge, etc.) are indistinguishable or \emph{permutable} (as Saunders calls them); this permutability expresses that labels do not carry physical meaning over and above the physical properties of the particles. For example, two widely separated electrons, following distinct trajectories and labeled $1$ and $2$, still represent exactly the same physical situation after we have exchanged labels (in the sense that the electron at the first position is now called electron $2$, and \emph{vice versa}). But this example also immediately demonstrates the physical insignificance of this type of indistinguishability: in spite of the exchangeability of the \emph{labels}, the \emph{particles} themselves can clearly be distinguished physically---we could easily tell them apart by a detection device that is sensitive to positions. So it seems that this permutability argument, and division by $N!$ on its basis, only relates to our mode of description and does not possess physical import. This division by $N!$ cannot possibly lead to different predictions; which makes it highly implausible that it could be relevant for a solution of the Gibbs paradox. We shall return to this in more detail in a moment, after having discussed another explanation for $1/N!$.

This alternative approach looks more promising from a physical point of view because it starts from a consideration of  \emph{open} systems. It can thus take into account the argument that as long as one stays with Eq.\ (\ref{boltzmann2}) and applies it to closed systems, it is impossible to justify any functional relation between variations in $S$ and variations in $N$ at all (\cite[\S 9]{ehrenfest}). We should switch to processes in which $N$ may \emph{change} (\cite{ehrenfest,vankampen}) in order to study entropy changes that are due to changes in $N$.

For this purpose we consider an ideal gas in a volume $V_1$ that is brought into contact with an ideal gas of the same kind in a volume $V_2$. If we assume the kind of randomness that is at the basis of statistical mechanics, the probability of having $N_1$ particles in subsystem 1 and $N_2 = N - N_1$ particles in subsystem 2 (with $N$ the fixed total number of particles) will be given by the binomial distribution
\begin{equation}\label{binom}
    P(N_1,N_2)= \frac{N!}{N_1!N_2!}(\frac{V_1}{V})^{N_{1}}(\frac{V_2}{V})^{N_{2}}.
\end{equation}

The particle number $N_1$ is now no longer a constant but has become a stochastic variable, which makes it possible to consider the change in probability and, by taking the logarithm, the change in entropy when $N_1$ varies. For example, if we let $N \rightarrow \infty$, $V_2 \rightarrow \infty$, while $N/(V_1 + V_2)$ is finite and constant, we find for the change of entropy of a system in contact with an infinite reservoir of particles:
\begin{equation} \label{entropybin}
\triangle S = \triangle \ln \frac{1}{N_{1}!}(\frac{V_1}{V})^{N_1}.
\end{equation}
There thus appears a factor $1/N_{1}!$ in the probability and entropy, and this time this factor is \emph{derived} instead of posited. The factor is due---via the binomial coefficient in the probability distribution---to the multiplicity of ways in which $N_1$ atoms can be chosen from a total of $N$ atoms. For this multiplicity to make sense it must be assumed that on the microscopic level of statistical mechanics the atoms are \emph{distinguishable}: it must make a difference \emph{which set} of atoms is selected from the total number of atoms $N$.

A typical example of application of Eq.\ (\ref{binom}) is the case in which a volume $V_1$ with $N$ ideal gas atoms is brought into contact with a larger volume $V_2$ that is empty. As soon as the partition is taken away, a situation arises which is not a thermodynamical equilibrium: since the probability (\ref{binom}) is very sharply peaked around $N_1 = N V_1/(V_1 + V_2)$, as a result of random motion a new equilibrium will be established in which the gas is (approximately) uniformly distributed over the total volume $V_1 + V_2$ (this is the essence of the second law in its probabilistic version). If $V_1 = V_2$, the associated change in entropy is, from Eq.\ (\ref{boltzmann}):
\begin{equation} \label{gibbs3}
\triangle S = k \ln \frac{N!}{(N/2)!^2} 2^{-N} = kN\ln 2.
\end{equation}

If we start with a volume $V_1$ filled with $N_A$ atoms of ideal gas $A$, and a volume $V_2 = V_1$ with $N_B = N_A$ atoms of ideal gas $B$, we find, analogously, for the entropy change when a partition between the gases is removed: $2kN_A \ln 2 = 2kN_B \ln 2$. This is exactly the entropy of mixing (\ref{mixing}) that thermodynamics predicts for this case.

If, however, we begin with two volumes with atoms of the \emph{same} gas, the situation immediately after the removal of the partition is already a state of maximum probability according to Eq.\ (\ref{binom}). In this case there will consequently be no increase in entropy, in accordance with the thermodynamical result. So Eq.\ (\ref{binom}), with its factors $1/N!$, appears to take away our Gibbs worries.

\section{Do the justifications for $1/N!$ really solve the Gibbs paradox?}

Although it is a frequent claim in the literature that division of the entropy expression by $N!$ because of fundamental indistinguishability or permutability  of particles (the first justification reviewed above) justifies discarding the Gibbs paradox, it is not difficult to see that this is incorrect. The reason is that transition to the reduced phase space, so that all insignificant permutations among particles of the same kind are factored out, should involve a division by $N!$ in which $N$ stands for \emph{all} particles of the same kind. For example, in the initial unmixed Gibbs situation, with the partition in place and with atoms of the same kind on both sides, $N$ should refer to the total number of atoms \emph{on both sides} of the partition: it clearly does not make any physical difference when we exchange labels across the partition.\footnote{Actually, we should in principle divide by the number of all particles of the same kind in the whole universe---this does not make a difference for our point about the irrelevance of this division made in the text. Compare also the later discussion of $1/N!$ in quantum mechanics.} But this implies that we correct the multiplicity of states by the \emph{same} numerical factor, both before and after the removal of the partition. All entropy \emph{changes} will therefore be exactly the same as when we do not make these corrections at all. In other words, division by $N!$ on the basis of fundamental indistinguishability or permutability of particles of the same kind only affects the absolute value of the entropy, of which we have already seen that it is without empirical significance. This division is not sensitive to changes in the situation like the removal of a partition or other mechanisms of mixing---it is therefore completely irrelevant to the Gibbs paradox.

The justification of $1/N!$ via making $N$ variable, and applying the binomial distribution, certainly looks more promising in this respect: here the division is not by a constant, but by the variable number of particles in a given volume. The idea of this derivation of $1/N!$ was first proposed by Ehrenfest and Trkal (\cite{ehrenfest}), in the early 1920-s. It is not as well known as it should be (but see the exposition by van Kampen (\cite{vankampen}); Swendsen has recently proposed a similar approach (\cite{swendsen})). Authors who comment on this justification for the occurrence of $1/N!$ in statistical mechanics expressions often create the impression that it provides a definitive solution of the Gibbs paradox within the framework of classical physics, by deriving the division by $1/N!$ from standard probability theory.\footnote{Saunders (\cite{saundersindist}) says that \emph{both} the permutability argument \emph{and} the Ehrenfest-Trkal argument provide a solution for the classical Gibbs paradox, and that quantum indistinguishability does the same for quantum particles. Our conclusion here will be that all three claims are wrong.} However, as we shall argue, this claim rests on a misinterpretation of the significance of the factor $1/N!$ that is thus derived.

The examples at the end of the previous section of the application of the Ehrenfest-Trkal method to the mixing of gases already suffice to make the essence of the situation clear. If two \emph{unequal} gases are allowed to mix, there is no equilibrium immediately after the removal of the partition, because the probability of the particle number is not at its maximum value. This means that a more probable situation will be established, in accordance with the second law in its statistical version. However, in the case of two \emph{equal} gases (at the same pressure and temperature) there \emph{is} an equilibrium, and nothing will change. It is this difference that is responsible for the presence or absence of an entropy of mixing in the examples. Where does this difference come from? Not from the binomial distribution or the $1/N!$ in it, but from our decision to look only at the number of particles of a particular chemical kind, without asking the question from where these particles came (from the left or right compartment). In the case of two chemically equal gases we decide not to follow the atoms from the left compartment, so that we renounce the possibility to identify them as different from atoms coming from the right; but this identification is exactly what we took for granted in the case of unequal gases (in which case the task is easy because of the chemical difference). There is a good reason for not bothering about the origin of chemically identical particles: the remaining differences are undetectable by the methods allowed in thermodynamics. As we discussed in the context of the Gibbs paradox in thermodynamics, we would leave the domain of thermodynamics if we were going to follow the individual trajectories of particles in order to detect the effects of mixing chemically equal gases. But this reason for the disappearance of the mixing entropy has nothing to do with the appearance of $1/N!$ in the binomial distribution.

Indeed, exactly the same situation arises when we work with the formula $S = k\ln W$ \emph{without} taking into account the $1/N!$ factor. As we know, in the case of unequal gases we find that there is an entropy of mixing, due to the increase in possibilities for the individual atoms of gases $A$ and $B$. After the removal of the partition each atom now has twice the number of available microstates it had before, which leads to an additional entropy of $S_{mix} = k \ln 2^{2N} = 2Nk\ln 2$. In the case of equal gases we find this same value \emph{if we still consider all microstates as different}---this is what leads to the Gibbs paradox in its second version. But if we pause and think about what is relevant to thermodynamics, we recognize that exchanges of atoms between the left and right compartments, although physically real, no longer lead to \emph{thermodynamically detectable} differences in the case of chemically equal gases. So we can safely diminish the total number of states, dividing by $2^{2N}$ in our calculation---not because there is no additional phase volume made available by taking the partition out, but because this additional phase volume cannot be made visible and usable by means of thermodynamical techniques. Fundamental `ontological' identity and indistinguishability do not play a role here.

Therefore, the Ehrenfest-Trkal binomial factor does not do any work in the solution of the Gibbs paradox. The crucial step that is responsible for the disappearance of $S_{mix}$ is the assumption that in the case of the mixing of two gases of the same kind we are not going to distinguish between atoms coming from $V_1$ and atoms coming from $V_2$. In mathematical terms, the disappearance of the mixing entropy is not due to the introduction of the coefficient $N_{A}! N_{B}!/N_{A1}! N_{A2}!N_{B1}! N_{B2}!$ in the formula for the probability of the distribution of the particles of sorts $A$ and $B$ over the compartments $1$ and $2$, but rather to \emph{our decision} to replace this coefficient by $(2N)!/(N!)^2$ in the case of gases $A$ and $B$ of the same kind. From the point of view of fundamental physics or mathematics there is nothing that compels us to take this step: the atoms from $V_1$ and $V_2$ are distinguishable entities and $(N!)^2/(N/2)!^2$ (the binomial factor appearing in the initial entropy for unequal gases if $N_A = N_B = N$) is simply numerically unequal to $(2N)!/(N!)^2$ (the factor for equal gases). What has happened here is that \emph{we} have decided to discard exchanges of atoms between the two volumes. In the case of unequal gases these exchanges \emph{would} count as changes of the total thermodynamical state---but if the gases are of the same kind they are considered irrelevant. So everything hinges on what we define as the thermodynamical state or changes in this state---and not on fundamental microscopic considerations about distinguishability and identity.

What, then, is the background of the idea that it is an undeniable fact that the Ehrenfest-Trkal derivation of $1/N!$ solves the Gibbs paradox? This comes from the misconception that the Ehrenfest-Trkal argument shows that $W$, in the formula $S = k log W $, should in all circumstances be divided by $N!$---as we have seen in section 3, this division would indeed remove the entropy of mixing, as a mathematical fact. But for this we need to redefine not only the entropies of the two mixing gas volumes, but also the entropy of the total system. Going back to the example at the end of section 3, to define away the entropy of mixing it is necessary that the entropy of the \emph{total} system, consisting of $2N$ particles, is divided by $(2N!)$. But the Ehrenfest-Trkal argument did not imply anything about the $N$-dependence of this total entropy. Indeed, the total number of particles is \emph{constant} in the Ehrenfest-Trkal argument, and not subject to any probability considerations. So there is no justification here at all for dividing the total number of states by $2N!$.\footnote{Swendsen (\cite{swendsen}) attempts to avoid this conclusion by an appeal to counterfactual reasoning: \emph{if} the total system were to enter into contact with another system, its number of particles would become variable and subject to a probability distribution. But such counterfactual reasoning has no explanatory force for the actual case of the Gibbs paradox, in which the total system is closed.}

So we see that the justification of $1/N!$ as proposed by Ehrenfest and Trkal does nothing to solve the Gibbs paradox\footnote{This is not to deny that the Ehrenfest-Trkal approach constitutes an essential step forward in the context of problems in which particle numbers can change, for example the study of dissociation equilibria!}; the conceptual situation remains exactly as it was.

\section{Quantum mechanics}

It is a central principle of quantum mechanics that a many-particles state of `identical particles' (particles of the same kind) must be completely symmetric or anti-symmetric under permutations of particle indices. This implies that the state does not change under permutations (except for a minus sign in the case of fermions), so that all probabilities and expectation values remain the same. This is what is meant by particle indistinguishability in the context of quantum mechanics: permutations of labels do not lead to a new physical situation.

At first sight this symmetry property of quantum states appears important for the Gibbs paradox, because it provides us with factors $1/N!$ at places that seem exactly right.\footnote{The factors appearing in quantum statistics are actually a bit more complicated. Suppose we are considering a system of $N$ particles, with $X$ one-particle states available to them. Classical counting leads to $X^N$ possible states for the $N$-particle system, if we assume that each one-particle state can be occupied by more than one particle. The number of quantum states for bosons is $(N + X - 1)!/N!(X-1)!$, which reduces to $X^N/N!$ if $X \gg N$ but in any case is a constant as long as $N$ does not change---this is the essential point for the argument in the text. In the case of fermions the number of states is $X!/N!(X-N)!$, which also reduces to $X^N/N!$ if $X \gg N$ and is also constant when $N$ does not vary.} Instead of $N!$ distinct product states, differing from each other by permutations of particle labels, we now have only one symmetrical or anti-symmetrical state.

But this reasoning based on counting of quantum states is spurious. To see why, note first that in the case of two different gases the expressions for the thermodynamic probabilities are multiplied by $1/N_{A}!$ and $1/N_{B}!$---but this is a multiplication by \emph{constants}, which only leads to the addition of a constant to the entropy. All entropy differences remain the same in this case; in particular, the entropy of mixing is exactly the same as in calculations without consideration of the symmetrization postulates. In the case of gases of the same kind, the factor that comes from the quantum counting is $1/(2N)!$ (for simplicity we again consider the case with equal amounts of gas, with a total number of $2N$ particles). But as before, multiplication of $W$ by this constant value only leads to the addition of a constant to the entropy, and the entropy of mixing is unaffected. So the introduction of factors that come from counting the number of quantum states does not have consequences for the calculation of the entropy of mixing, and the Gibbs paradox remains exactly what it was. This argument exactly mirrors the argument from the previous section about the physical insignificance of `permutability' in classical mechanics.

What then, is the background of the wide-spread conviction that quantum theory automatically solves the Gibbs paradox? This belief comes from an incorrect application of the (anti-)symmetrization postulates. According to this flawed reasoning, the number of states in each of the two compartments before mixing, with the partition in place, is assumed to be given by $X^N/N!$, with $X$ the number of available one-particle states, whereas for the total number of states after mixing $(2X)^{2N}/(2N)!$ is taken. In other words, the symmetrization rule is first applied to each volume separately, and then, after the removal of the partition, to the complete volume as one whole. But this way of counting is in conflict with what the symmetrization postulates require! These postulates require (anti-)symmetrization of the \emph{total} state of identical particles, even when the partition is still present. So what made the entropy of mixing go away in the criticized argument was not the quantum counting of states, but the different ways of applying the symmetrization postulates before and after the removal of the partition. If we apply the symmetrization postulates correctly, it is immediately clear that they cannot have consequences for the entropy of mixing as long as the total number of particles of the same kind is constant.

The often made mistake that we just identified has its origin in the rule of thumb that we do not need to (anti)symmetrize if particle wave functions do not overlap. The quantum states on different sides of the partition do not overlap in very good approximation, and our classical intuition even tells us that they do not overlap at all (which is false because the wave functions have infinite tails). So it might appear that we do not need to symmetrize the total wave function of all the particles in the container, as long as the partition is in place.

But this rule of thumb is indeed precisely a rule of thumb, a pragmatic rule justified by the fact that for the expectation values of \emph{certain} (namely local) observables it does not make a difference whether we symmetrize or not in the case of non-overlapping wave-packets. From a more fundamental point of view symmetrization is always necessary, and any failure to symmetrize the complete state would in principle show up in wrong predictions if expectation vales of arbitrary many-particles observables are considered (see, e.g., \cite[Vol.\ II, sec. XIV.8]{messiah}), \cite[Vol.\ 2, pp.\ 1406-1408]{cohen}). Indeed, on second thought it would be miraculous if the insertion or removal of a partition between two volumes of particles of the same kind would have instantaneous consequences for the symmetry properties of the total wave function! Symmetrization should always involve the total number of particles of the same kind, and this makes symmetrization completely irrelevant to the Gibbs paradox.

\section{Identity, distinguishability and permutability}

In the thermodynamical Gibbs paradox the essential point proved to be whether we are actually able to distinguish the gases coming from two different compartments. If we possess semi-permeable membranes, or a similar device that responds differently to the two gases, we will be able to verify experimentally that there is a mixing entropy---and if we do not discriminate between the gases, we will not encounter this entropy. The latter statement is true even if in principle it \emph{is} possible to make a distinction; although in this case more refined instruments \emph{would} of course identify an entropy of mixing after all. As we argued, this remains true if the gases are chemically the same. In this case we have to invoke the help of \emph{very} refined instruments, or Maxwellian demons, and this admittedly leads us outside of thermodynamics. But the point of principle remains that the absence or presence of an entropy of mixing does not correspond to the presence or absence of fundamental identity or (in)distinguishability of the gases, but rather to the applicability of a discrimination criterion chosen by us.

The situation is not different in Ehrenfest's and Trkal's treatment of the statistical background of the $N$-dependence of the entropy. Also here, the mixing entropy is there if we take the gases as different, and use a factor $ N!/N_A!N_B!$ in the calculation of the probabilities, and it is not there if the gases are treated as being of the same kind, in which case we use a factor $N!/(N_A + N_B)! = N!/N! = 1$. Again, the formulas do not dictate which option to choose.

The question naturally arises of whether everything is a matter of our choice and Nature has no say in this at all. Isn't there some point at which it becomes \emph{impossible in principle} to distinguish $A$ and $B$ atoms, whatever tools we use? In this case we could say that there is no mixing, and no entropy of mixing, in a fully objective sense.

In classical physics we can always find out, in principle, from where a particle came because of the existence of trajectories. It is therefore plausible to look to quantum mechanics if we want to find indistinguishability in principle; moreover, folklore teaches us that identical quantum particles are really fundamentally indistinguishable. In the previous section we have seen what this actually means mathematically: the permutation of `particle indices' does not lead to a physically different many-particles state. Take, for example, the case of two fermions described by two mutually orthogonal one-particle states $|\psi \rangle_1$ and $ | \phi \rangle_2$. The product state is excluded by the anti-symmetrization postulate, which says that instead the total state should be
\begin{equation}\label{twoelectrons}
| \Psi \rangle = \frac{1}{\sqrt{2}}(| \phi \rangle_1 | \psi
\rangle_2 - |\psi \rangle_1 | \phi \rangle_2 ),
\end{equation}
so that permutation of the indices only entails a change of sign. Does this state tell us that we are dealing with particles that cannot be distinguished?

That depends on how we assume that particles are represented in Eq.\ (\ref{twoelectrons}). If we think that the \emph{labels} $1$ and $2$ refer to particles, as a kind of particle name, then we find (via partial tracing) that these particles are in exactly the same state and posses exactly the same properties in the total state (\ref{twoelectrons}). But that is not at all how Eq.\ (\ref{twoelectrons}) is normally interpreted in the practice of physics! The usual interpretation of  Eq.\ (\ref{twoelectrons}) is that it represents a situation in which we have one particle in state $|\psi \rangle_1$ and one in state $ | \phi \rangle_2$, respectively. In other words, in practice the concept of a particle is not connected to the indices in the formalism, and the associated partial trace states, but by the one-particle states occurring in the total state---this also gives the correct classical particle limit (see for more detail and argument about this line of thought \cite{dieks,diekslubb,lombardi}).

This distinction between different ways of implementing the concept of a particle makes a difference for how the Gibbs situation should be understood in quantum mechanics. Assume that $|\psi \rangle_1$ is a state that is localized in the left compartment in a Gibbs experiment, and $ | \phi \rangle_2$ a state localized in the right compartment. If we uncritically follow the tradition of thinking of the labels as directly referring to particles, we have to accept that particle $1$ is delocalized, spread out over both compartments---and that particle $2$ is in exactly the same predicament. In this case, both particles are in the same mixed state
$$W = \frac{1}{2} ( |\phi \rangle \langle \phi | + | \psi \rangle \langle \psi | ).$$
\emph{These} particles will be indistinguishable: there will be no instruments that can tell them apart and even Maxwellian demons will be powerless to discriminate between $1$ and $2$.

But as already mentioned, this interpretation of what a particle is fails to make contact with the practice of physics and the ordinary meaning of `particle'. In reality, Eq.\ (\ref{twoelectrons}) would typically be used to represent the situation in which there is one particle in the left compartment, and one in the right compartment---the symmetry of the state indicates that the label has no physical significance, but this does not imply that the difference between the one-particle states is physically insignificant. If we accept this alternative way of looking at the situation, in which particles are associated with one particle states rather than with indices, the situation with respect to the Gibbs paradox turns out to be not basically different from what it was in classical mechanics: a Maxwellian demon will be able to distinguish the state $|\psi \rangle_1$ from $|\psi \rangle_2$. Indeed, if these two states are localized in two different disjoint compartments, like in the initial situation of the Gibbs thought experiment, they will be orthogonal. Under unitary evolution (as in the case of an ideal gas) this orthogonality will be preserved in time, and ideal yes-no experiments can in principle always provide a definitive answer about the origin of the one-particle state we are dealing with. This can be exploited by a smart demon to build up an entropy of mixing, with the help of an analogue of a semi-permeable membrane. It is easy to see that the distinguishability that is vital here will remain the same if we are dealing with a $2N$-particles state, describing $N$ particles to the left and $N$ to the right (before the removal of a partition). It will be possible in this case to define mutually orthogonal $N$-particles states in the respective compartments, and these will remain orthogonal under unitary evolution.

In other words, if we think of particles in the way just sketched, the situation with respect to \emph{distinguishability in principle} is not fundamentally different in quantum mechanics from the one in classical mechanics: even if distinguishing trajectories are not clearly defined, there is still orthogonality of states and a corresponding distinguishability. The presence or absence of an entropy of mixing in this case only depends on whether we are actually going to discriminate.

In the end we can only agree with van Kampen's statement (\cite[p.\ 311]{vankampen}) that in the case of ideal gases the Gibbs paradox is not different in quantum mechanics than in classical mechanics.

\end{document}